\documentclass[aip,jcp,reprint,amsmath]{revtex4-1}

\usepackage{natmove}
\usepackage{graphicx}
\usepackage[update,prepend]{epstopdf}
\usepackage{xcolor}

\newcommand{\ave}[1]{{\left\langle #1 \right\rangle}}
\newcommand{\deriv}[2]{{\frac{{\rm d} #1}{{\rm d} #2}}}
\newcommand{\blambda}{{\boldsymbol{\lambda}}}

\begin{document}
\title{Inverse methods for design of soft materials}

\author{Zachary M. Sherman}
\affiliation{McKetta Department of Chemical Engineering, University of Texas at Austin, Austin, Texas 78712, USA}

\author{Michael P. Howard}
\affiliation{McKetta Department of Chemical Engineering, University of Texas at Austin, Austin, Texas 78712, USA}

\author{Beth A. Lindquist}
\affiliation{Theoretical Division, Los Alamos National Laboratory, Los Alamos, New Mexico 87545, USA}

\author{Ryan B. Jadrich}
\affiliation{Theoretical Division, Los Alamos National Laboratory, Los Alamos, New Mexico 87545, USA}
\affiliation{Center for Nonlinear Studies, Los Alamos National Laboratory, Los Alamos, New Mexico 87545, USA}

\author{Thomas M. Truskett}
\email{truskett@che.utexas.edu}
\affiliation{McKetta Department of Chemical Engineering, University of Texas at Austin, Austin, Texas 78712, USA}
\affiliation{Department of Physics, University of Texas at Austin, Austin, Texas 78712, USA}

\begin{abstract}
Functional soft materials, comprising colloidal and molecular building blocks that self-organize into complex structures as a result of their tunable interactions, enable a wide array of technological applications. Inverse methods provide systematic means for navigating their inherently high-dimensional design spaces to create materials with targeted properties. While multiple physically motivated inverse strategies have been successfully implemented \textit{in silico}, their translation to guiding experimental materials discovery has thus far been limited to a handful of proof-of-concept studies. In this Perspective, we discuss recent advances in inverse methods for design of soft materials that address two challenges: (1) methodological limitations that prevent such approaches from satisfying design constraints and (2) computational challenges that limit the size and complexity of systems that can be addressed. Strategies that leverage machine learning have proven particularly effective, including methods to discover order parameters that characterize complex structural motifs and schemes to efficiently compute macroscopic properties from the underlying structure.  We also highlight promising opportunities to improve the experimental realizability of materials designed computationally, including discovery of materials with functionality at multiple thermodynamic states, design of externally directed assembly protocols that are simple to implement in experiments, and strategies to improve the accuracy and computational efficiency of experimentally relevant models. 
\end{abstract}

\maketitle

\section{Introduction}
Soft materials with tailored properties have found application in a variety of technologies including waveguides in photonic circuits \cite{Bogaerts2002}, collectors for energy harvesting devices,\cite{Nakayama2008} membranes for energy storage cells,\cite{Kanamura2005, Darling2014} and tunable-rheology fluids in brake lines, artificial joints, and vibrational dampeners. \cite{Klingenberg2001} The use-inspired behaviors of these materials derive from the physicochemical properties of their constituent components as well as their internal spatial organization (i.e., structure). Because fabrication methods that enable top-down control of structure at the nanoscale can be prohibitively expensive and slow for industrial-scale manufacturing processes \cite{Girard2006}, bottom-up strategies based on self-assembling materials have been explored as promising alternatives. \cite{Whitesides2002}  Colloidal nanoparticles, polymers, and proteins can serve as powerful material building blocks for self-assembly because their mutual interactions, which help determine the favored equilibrium state of the system, can be systematically varied through, e.g., their size, shape, charge, composition/sequence, and surface functionalization \cite{Glotzer2007}, providing a rich design space. A key challenge is to determine which building blocks reliably self-assemble a material with a targeted structure or desired macroscopic properties.

\begin{figure*}[!ht]
\centering
\includegraphics[width=\textwidth]{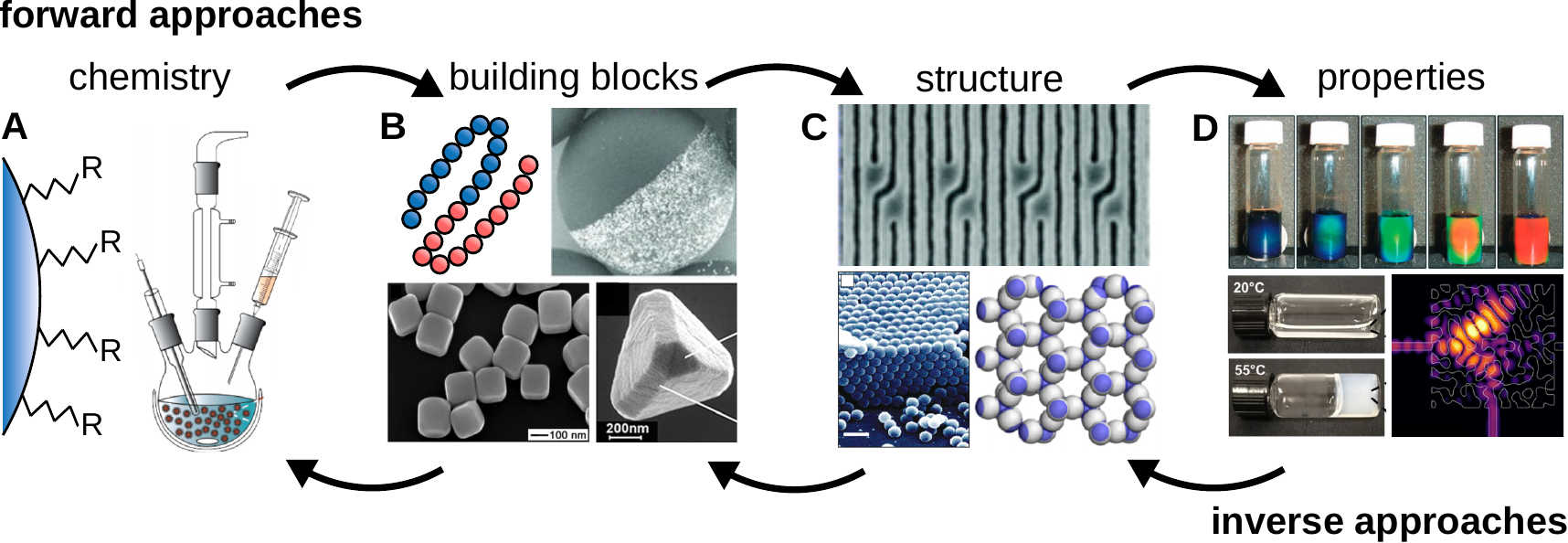}
\caption{Forward and inverse approaches in discovery and design of soft materials. In an example forward approach, materials with new properties may be discovered by repeatedly carrying out the following sequence of steps. Chemical synthesis (\textbf{A}) is used to create material building blocks with effective, coarse-grained interactions (\textbf{B}) that drive their assembly into structures (\textbf{C}) which impart emergent properties on the macroscopic scale (\textbf{D}). Inverse methods work backwards to systematically discover which material components will spontaneously form targeted structures or materials with desired macroscopic properties. For each panel from left to right and top down: \textbf{A:} Adapted with permission from \citenum{Agrawal2018}, Copyright 2018 American Chemical Society; \textbf{B}: Adapted with permission from \citenum{Loget2012}, Copyright 2012 Wiley; From \citenum{Sun2002}, adapted with permission from AAAS; From \citenum{Kalsin2006}, adapted with permission from AAAS; \textbf{C}: Adapted with permission from \citenum{Stoykovich2007}, Copyright 2007 American Chemical Society; Adapted with permission from \citenum{Jiang1999}, Copyright 1999 American Chemical Society; Adapted from \citenum{Reinhart2016}, with the permission of AIP Publishing; \textbf{D:} Used with permission from \citenum{Ge2007}, Copyright 2007 Wiley; Adapted with permission from \citenum{Cheng2019}, Copyright 2019 American Chemical Society; Adapted with permission from \citenum{Hughes2018}, Copyright 2018 American Chemical Society.}
\label{fig:design}
\end{figure*}

\emph{Forward} strategies for discovering new self-assembling materials are commonly adopted. In such approaches, an initial set of material building blocks is synthesized and protocols are chosen to promote their self-assembly in an experiment or a computer simulation. The structure and properties of the resulting material are subsequently characterized. These steps are then repeated (typically many times) with different choices for the building blocks or protocols to screen for materials with superior attributes. To make materials discovery more systematic and amenable to meeting specified design constraints, it can be helpful to instead formulate this process as an \emph{inverse} problem (Fig.~\ref{fig:design}). For example, one can define a figure of merit (FOM) based on a desired structure or macroscopic property and then apply methods of constrained optimization to help navigate the multidimensional design space and determine which available building blocks, interactions, or protocols are most suitable for realizing a material.

Progress on statistical mechanical approaches to inverse problems for designing soft matter has been chronicled in recent reviews and perspective articles. Topics covered include the design of colloidal interactions to stabilize self-assembled target structures \cite{Torquato2009} (Fig.~\ref{fig:design}C $\to$ \ref{fig:design}B) and the discovery of structures optimal for realizing desired macroscopic properties across a range of soft materials \cite{Jain2014} (Fig.~\ref{fig:design}D $\to$ \ref{fig:design}C). Inverse methods have proven powerful for designing granular materials \cite{Jaeger2015, Jaeger2016, Murugan2019}, block copolymer assemblies \cite{Jaeger2016, Gadelrab2017, Murugan2019}, and bio-inspired materials \cite{Murugan2019}. Recent reviews \cite{Ferguson2018,Jackson2019} have highlighted inverse techniques that leverage machine learning (ML) to effectively process the high-dimensional data obtained from computer simulations of materials to analyze and design their novel structural and dynamic properties. Inverse methods have also been widely applied in related fields including the design of molecules and chemical reactions\cite{SanchezLengeling2018, GomezBombarelli2018, Coley2019A, Coley2019B}.  

In this Perspective, we explore recent advances in the use of inverse methods for computational soft-material design. We split the discussion into methods related to structure design in Section \ref{sec:structdesign} and macroscopic property design in Section \ref{sec:propdesign}.  For structure design, a major challenge is discovering a FOM that can (1) discriminate between the target structure and its competitors and (2) encourage spontaneous assembly of the target.  For property design, the FOM is often known in advance and related to the property of interest. The challenge is to find an efficient way to compute and optimize it. The latter can be carried out either directly by varying the possible building block interactions (Fig.~\ref{fig:design}D$\to$\ref{fig:design}B) or in two stages by first discovering optimal structure (Fig.~\ref{fig:design}D$\to$\ref{fig:design}C) and then using that information to determine optimal interactions (Fig.~\ref{fig:design}C$\to$\ref{fig:design}B).  Within each section, we compare inverse methods recently demonstrated to be successful for addressing these problems. Promising methods that utilize ML algorithms have been proposed for both structure and property design strategies, and we discuss how they may be effectively incorporated into inverse schemes. Despite the successes of \emph{in silico} structure and property design, inverse techniques are not routinely used to design materials in experiments, \cite{Wilken2015, Walsh2019, Aharoni2018, Man2013, Miskin2013, Miskin2014, Auzinger2018, Reid2019, Hannon2013, Hannon2013B, Qin2013}  and improving the experimental realizability of computational design remains an outstanding challenging.  In Section \ref{sec:future}, we outline some promising future areas for which inverse strategies may be particularly effective and useful for directing experimental materials design. 

\section{Design for structure} \label{sec:structdesign}

\begin{figure*}
\centering
\includegraphics[width=\textwidth]{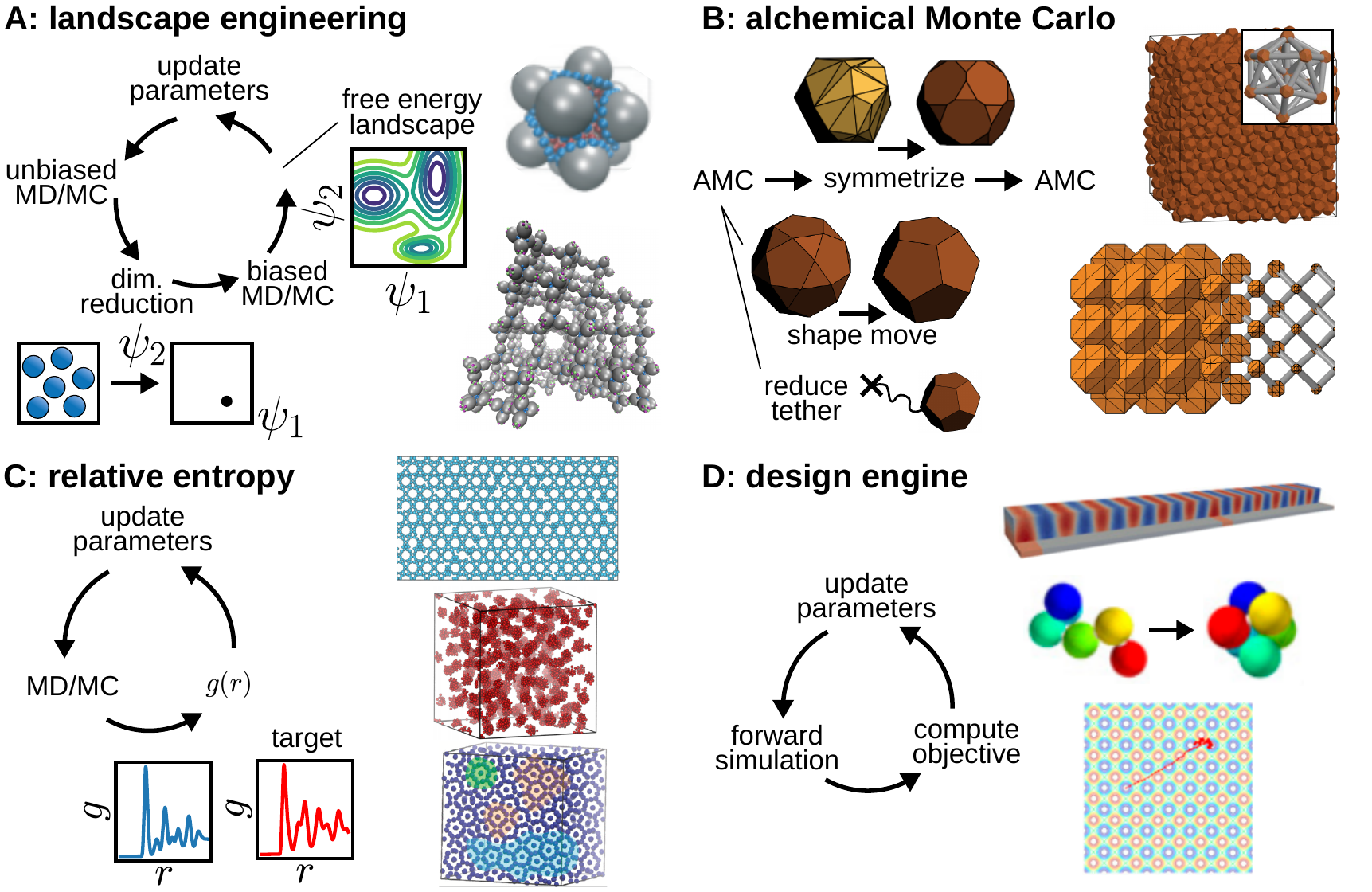}
\caption{Schematics illustrating the steps involved in several inverse methods along with snapshots of model materials designed and assembled using these approaches. \textbf{A:} Machine learning is used to discover structural order parameters $\psi_i$ from unbiased molecular dynamics (MD) or Monte Carlo (MC) simulations.  The free energy landscape is generated in the low-dimensional OP space using biased sampling methods, and the difference in free energy between the target and competitors is maximized.  This technique has been used to design patchy particles that assemble clusters\cite{Long2018}, which in turn assemble into open crystals.\cite{Ma2019}. Adapted from Refs. \citenum{Long2018} and \citenum{Ma2019} with permission from the Royal Society of Chemistry. \textbf{B:} Alchemical Monte Carlo (AMC) simulations find a particle shape that minimizes the alchemical free energy of a target lattice to favor its spontaneous assembly. Symmetrizing the shape can improve the target's thermodynamic stability.  This method has been used to assemble a rich variety of crystal structures from hard colloidal particles.\cite{vanAnders2015, Geng2019}. Modified from Ref. \citenum{Geng2019}. Copyright The Authors, some rights reserved; exclusive licensee American Association for the Advancement of Science. Distributed under a Creative Commons Attribution NonCommerical License 4.0 (CC BY-NC) \textbf{C:} The radial distribution function $g(r)$ from MD or MC simulations (blue) is compared to that for a target structure (red) to minimize the relative entropy between the two ensembles.  This method has been used to design isotropic pair potentials for assembling phases with complex morphologies.\cite{Jadrich2017, Lindquist2018} Adapted from Ref. \citenum{Jadrich2017}, with the permission of AIP Publishing. Adapted with permission from Ref. \citenum{Lindquist2018}, Copyright 2018 American Chemical Society. \textbf{D:} Forward simulations generate an ensemble of data that is used to skew the probability distribution toward configurations that contribute more toward a targeted structure or property than average.  This approach has been used to design block copolymers for templated self-assembly, folding polymers, and time-dependent processing conditions to shuttle a particle across an energy landscape. Adapted from Ref. \citenum{Miskin2016}.}
\label{fig:methods}
\end{figure*}

To design interactions that promote self-assembly into a specific structure (Fig.~\ref{fig:design}C$\to$\ref{fig:design}B), a target ensemble comprising the configuration data of the building blocks in the desired phase must be considered. Ideally, a single FOM can be constructed that is descriptive of the material's high-dimensional configurations and can be used to favor the target structure over those of competing phases. FOMs can include thermodynamic quantities, statistical distances from information theory, and structural order parameters. In this section, we discuss strategies that have used these types of FOMs to successfully design interactions for self-assembly of model materials into various target structures.  Some of these methods are depicted schematically in Fig. \ref{fig:methods}.

\subsection{Thermodynamic descriptors}\label{subsec:thermoFOM}

One of the pioneering approaches in computational design of self-assembly was to determine interactions that maximize the potential energy difference between a target structure and its competitors while ensuring mechanical stability of the target, effectively sculpting the ground-state potential energy landscape \cite{Rechtsman2005, Rechtsman2006, Rechtsman2007B, Torquato2009, Marcotte2011, Marcotte2011B, Marcotte2013}. In this strategy, the target structure and a pool of possible competitors are selected, and then the optimization is performed over a set of design parameters $\blambda$ characterizing the interaction potential between assembling components. While there are few inherent constraints on the functional form of the interactions, most research to date using this approach has focused on isotropic, pairwise potentials. A forward calculation of the ground-state phase diagram of the model with the optimized interactions can reveal if the list of the competitors considered needs to be expanded and further optimizations performed.

Using this method, interactions that stabilize several two-dimensional (e.g., honeycomb and square \cite{Rechtsman2005, Rechtsman2006, Marcotte2011, Marcotte2011B}) and three-dimensional (e.g. diamond \cite{Rechtsman2007B, Marcotte2013}) crystalline phases have been discovered. Modifications to this approach have enabled the optimization of interactions to ensure target stability over a wide range of particle concentrations. \cite{Jain2013, Jain2014B, Pineros2016, Pineros2016B, Pineros2017} Finite-temperature effects can also be treated approximately by minimizing the Lindemann criterion quantifying fluctuations from the target structure.\cite{Torquato2009, Rechtsman2005, Rechtsman2006, Rechtsman2007B}  While these methods are straightforward to implement for the design of interactions that stabilize crystalline targets, it is not clear how to extend them to target specific types of local structuring in disordered states of matter. Ensuring the kinetic accessibility of the target structure via self-assembly (e.g., from a disordered fluid) is also an outstanding challenge for ground-state-based strategies.  

Long and Ferguson recently developed a free-energy ``landscape engineering'' method (Fig. \ref{fig:methods}A) that goes beyond potential-energy minimization. In this approach, the free energy of a desired structural motif is directly minimized relative to other possible emergent structures in a low-dimensional space of collective coordinates. Importantly, the collective coordinates, or ``order parameters'' (OPs), are machine-learned from the high-dimensional space of raw particle coordinates to maximize information retention. Diffusion maps\cite{Ferguson2010, Long2014, Chiavazzo2017, Long2018} and autoencoders\cite{Chen2018C, Chen2018D, Ribeiro2018, Sultan2018} have been shown to be particularly useful for this reduction, but in principle other ML techniques or physically informed OPs (Section \ref{sec:orderparameters}) could be used. The free-energy landscape is computed in the low-dimensional OP space using enhanced sampling techniques \cite{Long2018,Chiavazzo2017, Chen2018C, Chen2018D}.  The free-energy differences between the target and its competitors are extracted from the landscape and used to update the design parameters in an iterative loop, depicted in Fig.~\ref{fig:methods}A. In contrast to potential-energy minimization methods, landscape engineering naturally incorporates temperature effects and automatically enumerates competitors from the free-energy landscape.  Landscape engineering has been used to construct patchy colloids that self-assemble into targeted clusters \cite{Long2018} which in turn assemble into open crystals \cite{Ma2019}, as seen in the snapshots in the right side of Fig.~\ref{fig:methods}A.

Constructing an entire free-energy landscape is difficult due to its computational costs, so landscape engineering has not yet been applied to more general target structures.  However, even if the landscape cannot be computed, it can still be navigated and updated in simulations.  Van Anders, Glotzer, and coworkers proposed a design approach using simulations in an ``alchemical'' (or ``expanded'') ensemble (Fig. \ref{fig:methods}B) \cite{vanAnders2015, Geng2019}.  Their work focused on a family of hard particles, the shapes of which are described by a set of variables, $\blambda$. For athermal particles, the free energy for a given shape is purely entropic with contributions from positions, $\mathbf{x}$, and orientations, $\mathbf{q}$. The particles were initially constrained to a target lattice using a tethering potential $E(\mathbf{x})$. Alchemical Monte Carlo simulations were performed in an ensemble where particles not only rotated and translated but also fluctuated in shape. The partition function $Z_{\rm AMC}$ of an ensemble of $N$ identical particles can be expressed as
\begin{equation}\label{eq:alchemical}
Z_{\rm AMC} = \int d\mathbf{x} \, d\mathbf{q}\, d\blambda \, e^{-\beta [U(\mathbf{x}, \mathbf{q} | \blambda) - N \boldsymbol{\mu}\cdot\blambda + E(\mathbf{x})]},
\end{equation} 
where $\boldsymbol{\mu}$ are alchemical potentials (conjugate to the shape parameters) and $U(\mathbf{x}, \mathbf{q} | \blambda)$ is the interparticle potential.

Because $\boldsymbol{\mu}$ cannot be controlled in real systems, $\boldsymbol{\mu}$ was set to zero to avoid biasing the particle shape, and the external tethering potential was slowly reduced. If the crystal structure is stable in the limit $E\rightarrow 0$, the free energy (entropy) for the target lattice has been minimized (maximized) with respect to particle shape.  As shown in Fig.~\ref{fig:methods}B, the particle shape can be additionally symmetrized to improve the thermodynamic stability of the target.\cite{Geng2019}  Because the free energy, rather than a free energy difference, is the FOM, alchemical Monte Carlo does not explicitly consider competitors. This is computationally efficient because it avoids having to fully sample the free-energy landscape but does not ensure that competitors are disfavored and thus the target may only be metastable. Nonetheless, this method has been used successfully to optimize hard-particle shapes that assemble into many complex crystals (see, e.g., right side of Fig.~\ref{fig:methods}B).\cite{Geng2019} Alchemical methods are not limited to hard particles and $\blambda$ may include parameters characterizing other particle interactions \cite{Zhou2019}.

\subsection{Statistical distances}
Information theory provides quantities, so called ``statistical distances'', that characterize differences between data samples.  The Kullback-Leibler (KL) divergence is one such quantity;~\cite{Barber2012,Chialvo2015} the KL divergence of a target distribution, $P_{\rm t}(\mathbf{x})$, with respect to a model probability distribution, $P(\mathbf{x} | \blambda)$, is defined as
\begin{equation} \label{eq:relent}
D_{\rm KL} (P_{\rm t}(\mathbf{x}) || P(\mathbf{x} | \blambda)) = \int d\mathbf{x} P_{\rm t}(\mathbf{x}) \ln\left[P_{\rm t}(\mathbf{x})/P(\mathbf{x}|\blambda)\right].
\end{equation}
In the case of equilibrium self-assembly, $P(\mathbf{x} | \blambda)$ is the Boltzmann weight of a configuration $\mathbf{x}$, and $\blambda$ are the parameters that characterize the potential energy function. By optimizing $\blambda$ so that $D_{\rm KL}$ is minimized, the structures self-assembled from the model system are made to resemble the target configurations. Minimization of $D_{\rm KL}$ is conceptually appealing and intuitive as a design objective because it is equivalent to maximizing the likelihood that the probabilistic model $P(\mathbf{x} | \blambda)$ will sample the configurations contained in the target ensemble.~\cite{Jadrich2017} 

While various computational approaches have been applied to minimize $D_{\rm KL}$, \cite{Shell2008, Chaimovich2011, Bilionis2013} updates to $\blambda$ consistent with a steepest-descent optimization are particularly simple to compute in the canonical ensemble when the interaction potential is pairwise and isotropic (denoted here as $u(r|\blambda)$). In particular, the potential parameters used in a molecular simulation during the $(i+1)^{th}$ iteration of the minimization, $\blambda_{i+1}$, are determined from those used in the $i^{th}$ iteration, $\blambda_{i}$, as
\begin{equation} \label{eq:relentiter}
\blambda_{i+1} = \blambda_i + \alpha \int d\mathbf{r} \, \Big( g(r | \blambda_i) - g_{\rm t}(r) \Big) \big[ \nabla_\blambda u(r | \blambda) \big]_{\blambda_i},
\end{equation}
where $ g(r | \blambda_i)$ and $g_{\rm t}(r)$ are the radial distribution functions of the simulated model in the $i^{th}$ iteration and that of the target ensemble, respectively, and $\alpha$ is a tunable parameter that controls the magnitude of the update. This iterative update process is depicted schematically in Fig.~\ref{fig:methods}C. Unlike the thermodynamic-descriptor-based methods discussed in Sect.~\ref{subsec:thermoFOM}, which are not guaranteed to result in self-assembly of the target in a forward simulation, Eq.~\ref{eq:relentiter} uses the structures measured from the self-assembly process as input to the parameter update at each optimization step. In this way, spontaneous assembly of the target structure is strongly promoted over its competitors by the interactions optimized using such ``on-the-fly'' methods.~\footnote{In practice, this necessitates that at each optimization step the system is started from a disordered state and not configurations from the previous step.}

The strategy described above has been used to design model isotropic pair potentials that self-assemble exotic structures (Fig.~\ref{fig:methods}C) including open lattices,\cite{Lindquist2016B, Jadrich2017} Frank-Kasper phases\cite{Lindquist2018}, multi-component crystals,\cite{Pineros2018} and colloidal strings.\cite{Banerjee2019}. Some desirable features of this approach include: (1) by manipulating the form of $u(r| \blambda )$, physically motivated interaction potentials can be discovered, \cite{Lindquist2016, Lindquist2019} (2) by varying the ensemble in which the iterative simulations are performed, simultaneous control of structure and thermodynamic quantities, such as the pressure of the self-assembled system, can be achieved, \cite{Lindquist2019} and (3) the minimization can also be performed in Fourier space, which may be computationally convenient for some design problems. \cite{Adorf2018}

The Kullback-Leibler divergence is also termed the relative entropy. Its minimization has been used to parameterize molecular coarse-grained models, where many atoms might be represented as a single bead, that are intended to stand in for more computationally expensive all-atom target simulations. \cite{Noid2013,Shell2008, Chaimovich2011} In both design for self-assembly and coarse-graining, the goal is to discover the parameters for a probabilistic model that are most likely to reproduce a target data set, whether that data set comes from an all-atom simulation or a contrived set of configurations that display a desired structural motif. Given these similarities, it is perhaps not surprising that other techniques from the coarse-graining literature have found success in design for self-assembly applications as well. For example, iterative Boltzmann inversion, which utilizes a heuristic update scheme with the same stationary point as Eq.~\ref{eq:relentiter} for a pair potential that is infinitely flexible, has been used to discover isotropic pair interactions that self-assemble cluster fluids\cite{Jadrich2015} as well as mesoporous materials\cite{Lindquist2016, Lindquist2017}. 

The relationship between coarse-graining and design for self-assembly problems suggests multiple avenues for future work on the latter. For example, coarse-graining has been performed with multi-body\cite{Sanyal2016, Zhang2018B} and anisotropic\cite{Paramonov2008} interactions.  These more complex models are compatible with the relative entropy framework described above. Because many interactions commonly used to assemble structures in experiments are many-body and/or anisotropic in nature, including those mediated by electric charges,\cite{Kalsin2006} electric and magnetic fields,\cite{Yethiraj2003, Swan2012} surface tension,\cite{Yao2015, Sharifi-Mood2015} nematic liquid crystals\cite{Smalyukh2018}, and heterogeneous surfaces\cite{Chen2011, Wang2012}, embedding these features into the design space may allow for stronger coupling between computational and experimental materials assembly.   Additionally, certain target structures may require very complex interactions (or may even be impossible) to assemble if the potential is restricted to isotropic and pairwise forms. Nonetheless, there may be a ``simpler'' many-body and/or anisotropic potential that will readily assemble the structure. For example, the formation of capsid-like structures would undoubtedly be difficult for particles with isotropic interactions, but patchy particles with relatively simple short-ranged interactions are known to self-assemble into them with high fidelity.\cite{Long2018}. 

Finally, other statistical distances can also serve as FOMs. For example, one drawback of relative entropy minimization is that $D_{\rm KL}$ is not readily amenable to optimizing singular interactions such as a hard core-potential. A hard core produces regions in configuration space of zero weight (i.e., $P(\mathbf{x}|\blambda) = 0$) which leads to a divergent $D_{\rm KL}$ if $P_{\rm t}(\mathbf{x}) \ne 0$ for the same configurations.  In such cases, the relevant gradients cannot be computed to minimize $D_{\rm KL}$.  In contrast, the Bhattacharyya distance\cite{Chialvo2015}
\begin{equation} \label{eq:bhatta}
D_{\rm B}(P_{\rm t}(\mathbf{x}), P(\mathbf{x|\blambda})) = -\ln\left[ \int {\rm d}\mathbf{x} \sqrt{P(\mathbf{x|\blambda}) P_{\rm t}(\mathbf{x})} \right]
\end{equation}
does not share this limitation and might be used as an alternative metric for inverse design of hard-particle systems. 

\subsection{Structural order parameters} 
\label{sec:orderparameters}
\begin{figure*}
\centering
\includegraphics[width=0.9\textwidth]{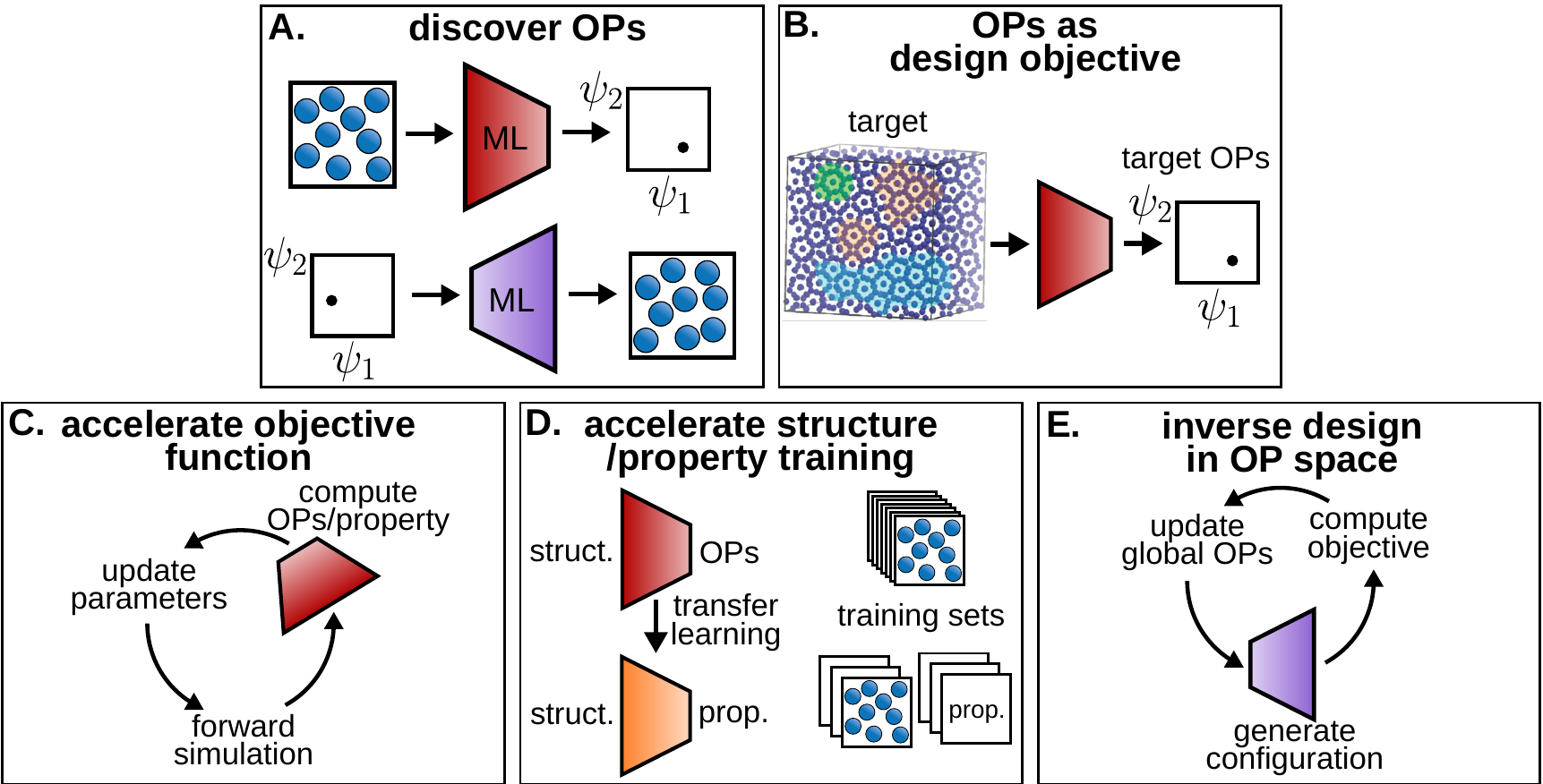}
\caption{Strategies for leveraging machine learning (ML) architectures for enhanced materials design.  \textbf{A:} Unsupervised training on a large training set of configurations discovers functions (represented with trapezoids) that map between high-dimensional configurations and a low-dimensional set of order parameters (OPs) $\psi_i$.  \textbf{B:} The ML OPs can be used as the design objective for assembling complex structures.  Adapted with permission from Ref. \citenum{Lindquist2018}, Copyright 2018 American Chemical Society. \textbf{C:} ML reduces the cost of computing objective functions in inverse design. D: The structure-OP function can be transferred to accelerate supervised learning of structure-property relations from a small training set. \textbf{E:} Inverse design can be performed directly in low-dimensional OP space using the ML configuration-generating function in lieu of forward simulations.}
\label{fig:transferlearning}
\end{figure*}

Many of the preceding descriptors are statistical mechanical quantities that must be computed on the basis of an ensemble of configurations. This limits their usefulness for systems where the relative statistical weights of configurations are not readily known (e.g., non-equilibrium systems). A more generically applicable strategy is to instead use a structural OP that serves as a low-resolution description of a high-dimensional configuration. When such OPs reliably distinguish between a target structure and its competitors,\cite{Keys2011B} they can be used to steer an iterative scheme using the OPs as the FOM, like in Figs.~\ref{fig:transferlearning}A--C. For example, Kumar \textit{et al.}~recently used the Steinhardt bond-order parameters based on spherical harmonics of local neighbor orientations\cite{Steinhardt1983, Lechner2008} to design pair interactions for assembly of body-centered-cubic colloidal crystals \cite{Kumar2019}. However, OPs that are sufficiently discriminatory between target structures and competitors can be challenging to construct, particularly for complex structures like open lattices, crystals with large unit cells, and quasicrystals as well as cases where potential competitor structures are not known in advance.  Machine learning offers possible solutions to automatically discover OPs from structural data for design.

Several supervised ML methodologies using neural networks have successfully classified input configurations according to a library of known structures.\cite{Geiger2013, DeFever2019, Fulford2019}  The ML classifiers outperform classifications using traditional OPs based on local orientations,\cite{Steinhardt1983, Lechner2008} angles,\cite{Chau1998, Errington2001} and neighbor-graph topology\cite{Larsen2016} in discriminating complex crystal phases\cite{Geiger2013, Fulford2019} and can be trained to identify interfacial structures.\cite{DeFever2019}  If relevant structures are not known ahead of time or are difficult to produce, unsupervised ML methods leveraging clustering algorithms can categorize similar structures together.\cite{Phillips2013, Spellings2018, Adorf2019}

These methodologies primarily classify configurations into discrete categories, but continuous OPs are desirable for optimization. These OPs can be local, for example, computing a descriptor for each particle, or global, computing one value for an entire configuration.  The spatial resolution of local descriptors makes them useful for characterizing interfaces; however, many conventional local order parameters (e.g., the Steinhardt parameters) are not constructed to accurately identify interfaces.  Therefore, Reinhart and coworkers developed an unsupervised ML method, called neighborhood graph analysis \cite{Reinhardt2017A, Reinhardt2017B, Reinhardt2018}, that uses diffusion maps to discern a few continuous OPs characterizing local structural motifs; their method efficiently discriminated between not only a variety of colloidal crystals but also their surfaces and defects.  Neighborhood graph analysis has also been useful for understanding properties of grain boundaries.\cite{Snow2019} 

Global OPs are useful to compute a single FOM to update parameters in an iterative design loop. One fruitful strategy for generating global OPs is to perform dimensionality reduction on a large data set of configurations and use the low-dimensional representation as an OP. While discovery of the OP requires multiple configurations, once defined, the OP can be computed on a per-configuration basis. Dimensionality reduction methods such as diffusion maps \cite{Ferguson2010, Long2014, Chiavazzo2017, Long2018}, autoencoders \cite{Chen2018C, Chen2018D, Ribeiro2018}, and variational dynamics encoders \cite{Sultan2018} have been used to construct global OPs that are continuous and differentiable. The underlying principle of such dimensionality reduction approaches is to find an intermediate compressed representation that when uncompressed is as close to the input data as possible, as depicted in Fig. \ref{fig:transferlearning}A. Such OPs have been leveraged for enhanced sampling of molecular dynamics trajectories directly in the low-dimensional OP space.\cite{Long2014, Chiavazzo2017, Long2018, Chen2018C, Chen2018D, Ribeiro2018, Sultan2018, Yang2018B} Similarly, Jadrich \textit{et al.}~used sorted arrays of pairwise distances and orientations to obtain global OPs through principal component analysis (PCA). \cite{Jadrich2018A, Jadrich2018B} The resulting OPs were able to detect a variety of phase transitions including freezing in hard disks/spheres, liquid-gas and compositional phase separation, nematic ordering in ellipses, and a non-equilibrium phase transition.  A similar method utilized nonnegative matrix factorization to compute global OPs in a ternary lipid mixture.\cite{Lopez2019} 

For structural design, the machine-learned OPs of a target can serve as a convenient, numerical design objective, as shown in Fig.~\ref{fig:transferlearning}B. The output OPs can be employed in an iterative scheme (Fig.~\ref{fig:transferlearning}C) whereby configurations are converted to OPs and parameters can be tuned to push the OPs towards the desired value. This may be especially useful to design complex structures for which good descriptors are lacking.  However, it is not necessarily straightforward to perform such an optimization.  Learning the OPs requires generating a large amount of data.  This data must be representative of the structures likely to be sampled during design so that the OPs can discriminate between the target structure and competitors.  For machine-learned OPs to be reliably ported into inverse schemes, it would be beneficial to have systematic methods for determining the minimimum amount of data required to learn sufficiently accurate OPs, where in the design space this data should be collected, and if the data can be acquired on-the-fly and/or recycled between designs using transfer learning.  Similarly, automated ways to select the best FOM, ML strategy, and iterative scheme for a certain design problem would be useful for non-experts.  Such methods have not been fully explored for structural design problems and remain an important area for future research.

\section{Design for properties} \label{sec:propdesign}
Unlike the design of interactions for self-assembling a target structure, where determining a suitable FOM was challenging, there is an obvious choice for a FOM in property design---the property itself. Each iteration of the optimization involves computing the property from the underlying structure, i.e., evaluating the ``structure--property relationship''.  If the structure--property relationship is known or readily computed, material properties can be designed using a variety of optimization routines. For example, Miskin and Jaeger designed an unusual strain-stiffening granular material using an evolutionary optimization algorithm \cite{Miskin2013, Miskin2014}, while elastic networks with maximally negative Poisson ratios \cite{Reid2019} and targeted allosteric response \cite{Yan2017} have been designed using gradient-descent and simulated-annealing algorithms, respectively. Dynamic properties like diffusivity and viscosity can be optimized by similar techniques \cite{Goel2008, Carmer2012, Monroe2018}.

More often though, material properties are complex functions of structure that can depend on dynamic or nonequilibrium behavior, and the structure--property relationship is prohibitively expensive to evaluate frequently. In this case, either (1) the iterative optimization algorithm must be significantly improved to reduce the total number of structure--property evaluations required for convergence or (2) the cost of computing the property must be reduced by evaluating the structure--property relationship approximately.

\subsection{Iterative schemes}
Many techniques have been developed to improve optimization routines relevant for design of soft matter, and it is beyond the scope of this Perspective to comprehensively cover them.  Inverse methods leveraging Bayesian optimization appear particularly promising and have been recently adapted for design of material properties.\cite{Yang2018B, Ju2017, Tran2019}  In addition to navigating design spaces efficiently, these methods also provide estimates for uncertainties and sensitivities of solutions, which may be useful for finding and prioritizing degenerate solutions.  For example, the solution that is least sensitive to perturbations in the design parameters of a model might be the best to fabricate in experiments, where deviations from the model are bound to occur.

We highlight one particular approach to improve the convergence of optimizations that is physically motivated and has been applied to property design of self-assembled materials.   Jaeger, de Pablo, and coworkers proposed a statistical physics ``design engine'' \cite{Miskin2016}, depicted in Fig.~\ref{fig:methods}D,
\begin{equation} \label{eq:designengine}
\deriv{P(\mathbf{x}|\blambda)}{t} =  P(\mathbf{x}|\blambda) \left(f(\mathbf{x}) - \ave{f(\mathbf{x})}_{P(\mathbf{x}|\blambda)}  \right),
\end{equation}
that prescribes dynamics to the optimization with an artificial time $t$. Here, $f$ is an objective function of the configuration $\mathbf{x}$ and sets the design goal, and $\ave{\cdot}_P$ is an ensemble average over the probability distribution $P(\mathbf{x}|\blambda)$. The design engine leverages information about the entire probability distribution; configurations that contribute more to $f(\mathbf{x})$ than average increase their likeliness, while those that contribute less than average become more unfavorable.  The form of Eq.~\ref{eq:designengine} enforces conservation of probability
and ensures the probability distribution is normalized.  The design engine converges more quickly than standard optimizers (like steepest descent or simulated annealing) for certain classes of problems. There is considerable flexibility in choosing $f$ so that various OPs or materials properties can be incorporated for either structure or property design. The design engine has been successfully applied to a sampling of inverse problems shown in the right side of Fig.~\ref{fig:methods}D, including colloidal crystallization,\cite{Kumar2019} polymer folding, self-assembly of block copolymers, and even nonequilibrium systems.\cite{Miskin2016}  Other types of physically motivated iterative schemes may also be useful; for example, alchemical-ensemble methods have been suggested for property design. \cite{Zhou2019}  

\subsection{Machine learning for accelerated property design} \label{sec:structprop}
ML has proven effective for reducing the cost of determining material properties (Fig. \ref{fig:transferlearning}C--D).  One ML strategy is to discover an easier-to-compute OP that serves as a proxy for difficult-to-compute properties.  Support vector machines have been used to analyze the ``softness'' of glassy systems from their structural features \cite{Schoenholz2016, Schoenholz2017, Cubuk2017}. PCA of particle configurations has been used to find OPs for the mechanical properties of polycrystalline materials\cite{Paulson2017} and the effective diffusivity through membranes \cite{Cecen2014}, and the OPs were then regressed to simulated material properties. More accurate predictions can be obtained by using supervised ML methods to learn the structure--property relationship directly.  Neural networks were trained to predict the elastic modulus of a lattice model of a binary elastic composite from its configuration, outperforming linear regression of OPs from PCA.\cite{Yang2018}  Neural networks similarly outperformed regression to predict the activity of antifreeze proteins from the structure and hydrogen-bonding dynamics of nearby water.\cite{Kozuch2018}

Learning these relationships can require large training sets, which are impractical to generate if the material property is difficult to compute in the first place.  Transfer-learning strategies can be used to accelerate training.  Yang, Agrawal, and coworkers trained a generative adversarial network (GAN) for heterogeneous, disordered two-dimensional optical materials.\cite{Yang2018B} The GAN was initially trained on a large data set of configurations that were easy to produce; this learning was leveraged to initialize a new network for computing optical adorption using a smaller, more expensive-to-produce training set of configuration--adsorption pairs (Fig. \ref{fig:transferlearning}D).  The transfer-learned structure--property network is more accurate than a network trained from scratch for a fixed number of iterations and training set size, or equivalently, requires smaller training sets and fewer iterations to achieve the same prediction accuracy. Transfer-learning may also be useful in cases where a design optimization pushes the target property outside the bounds of the training set so that retraining of the ML structure--property relationship is required, but such methods have yet to be fleshed out for materials design. 

During training, generative ML methods learn a small set of OPs from which they are able to generate new configurations statistically indistinguishable from those in the training set. The generator from a GAN was recently used to perform inverse design directly in the OP space to find structures with high optical absorption (Fig. \ref{fig:transferlearning}E).\cite{Yang2018B}  Because OP space is much lower-dimensional than configuration space, it is easier to explore. Other ML approaches for dimensionality reduction can be used similarly for inverse design; for example, Guo, Ren, and coworkers designed density (spatial) distributions of heat-transfer materials with optimal thermal properties using the decoder from a pretrained autoencoder network to perform the optimization in OP space.\cite{Guo2017}  Similarly, structural OPs discovered using the ML methods of Sec.~\ref{sec:orderparameters} can serve as the design space. Combining both ML OP design spaces and ML structure--property relationships in the same inverse cycle could provide even more efficient design schemes.

\subsection{Multiobjective design}
The highlighted inverse methods are primarily intended for design of a single material property, but many applications require materials with multiple functionalities. For example, bulletproof vests should be lightweight and flexible yet highly energy-dissipative,\cite{Jaeger2015} while membranes used in flow batteries must be both mechanically strong and electrically conductive. \cite{Kanamura2005}. Methods that can efficiently address inverse problems with several design objectives will be useful for such state-of-the-art materials.

One possible strategy is to reduce these ``multiobjective'' design problems into a single objective that depends on multiple material properties. This idea was used to engineer several mechanical properties of a gallium-iron alloy \cite{Liu2015} and to design multifunctional optical ports.\cite{Hughes2018} However, there are many ways to incorporate several criteria into a single objective function, and the arbitrary choice has a large effect on the final solution.\cite{Liu2015, Hughes2018}  Such design problems may be better approached using algorithms developed for multiobjective optimization,\cite{Laumanns2002, Marler2004} but this application has not been thoroughly explored for soft-materials design.

\section{Future directions} \label{sec:future}
The methods summarized in this article have been remarkably successful for designing soft materials \emph{in silico}.  Inverse approaches have similarly been used to design materials in experiments,\cite{Wilken2015, Walsh2019, Aharoni2018, Man2013, Miskin2013, Miskin2014, Auzinger2018, Reid2019, Hannon2013, Hannon2013B, Qin2013} but these strategies have not taken advantage of the methods developed for \emph{in silico} design.  As a result, there are compelling future opportunities to address the translation of effective computational strategies for the discovery of new materials to the laboratory. These opportunities include the application of inverse approaches to find robust solutions subject to experimentally realistic design constraints (Sec.~\ref{sec:multistate}), the adaptation of design assembly protocols that are simple to implement in experiments (Sec.~\ref{sec:externalfields}), and the development of strategies to improve the accuracy and computational efficiency of experimentally relevant models (Sec.~\ref{sec:experimentconnect}).

\subsection{Multistate design}
\label{sec:multistate}
Most inverse methods for structure design are intended for a single target structure at one thermodynamic state point (e.g., temperature and pressure).  This is problematic in practice because processing and operating conditions are rarely constant over a material's lifetime.  Materials designed only for one state may have different structures and properties that are suboptimal or even unusable at other conditions.  Alternatively, a material with a structural transition may be the design objective.  For example, reconfigurable materials that change their structure in response to their conditions are useful for sensing applications\cite{Holtz1997, Lee2000} and as responsive materials capable of controlled, on-the-fly modulation of properties.\cite{Klingenberg2001, Ge2011}  Methods allowing for design of multiple target structures and multiple state points can efficiently address these inverse problems.

The coarse-graining community has addressed a problem closely related to ``multistate'' design: developing an optimal coarse-grained representation from atomistic data sampled at different thermodynamic states.\cite{Mullinax2009, Moore2014, Sanyal2019, Sharp2019} Such approaches could be leveraged for inverse schemes to find a single interaction potential that assembles different structures under different conditions. In principle, entire phase diagrams could be designed by tessellating state points with target structures and simultaneously designing for them. This approach could systematically find materials with exotic phase behavior like those that ``inverse melt'' upon cooling. \cite{Feeney2003} However, computational demands for this procedure may be intense, and a feasible solution that assembles all target structures may not exist. Investigation is needed to demonstrate the possibilities for and limitations of multistate design.

In addition to equilibrium thermodynamic considerations, reconfigurable materials require kinetic transitions from one structure to another. Particularly challenging are fluid--solid and solid--solid transitions, which have many kinetic barriers but are essential for controlled manipulation of material properties.\cite{Grzelczak2010, Sherman2019} Objective functions that vary during the optimization may help find solutions with transitions specifically embedded. For example, an objective function that periodically switched between two states of reconfigurable circuits \cite{Kashtan2005} and of allosteric networks \cite{Hemery2015} found solutions prioritizing transitions between states. Murugan and Jaeger recently suggested applying this same ``switching'' strategy to self-assembling materials \cite{Murugan2019}, and we agree that this is a potentially fruitful area for further study.

\subsection{Engineering assembly protocols with external controls}
\label{sec:externalfields}
If interactions among building blocks can be controlled with an external stimulus, the stimulus may be used to facilitate self-assembly.  Many such systems have been studied experimentally, including materials that respond to light, temperature, electric/magnetic field, and flow.\cite{Grzelczak2010} This approach for assembly is attractive because it is often easier to control and modulate the external processing conditions than it is to change the physicochemical properties of the building blocks.

If the interactions induced by the external stimuli can be represented with simple expressions, for example in terms of an equilibrium interparticle pair potential, they are amenable to design using inverse methods such as those in Sections~\ref{sec:structdesign} and \ref{sec:propdesign}.  Often though, these interactions are many-bodied (\emph{e.g.} electrostatic), anisotropic (\emph{e.g.} dipolar), and out-of-equilibrium (\emph{e.g.} flow-induced).  This complicates design of the parameters of externally induced interactions, but if robust methods for this inverse problem could be developed, the solutions may be easier to realize experimentally.

A particularly promising feature of induced interactions is the ability to vary them over time.  For example, these approaches show promise for enhancing crystallization rates while reducing defects\cite{Tang2016, Tang2017, Swan2012, Howard2018} as well as assembling structures not stable at equilibrium.\cite{Sherman2019} Processing conditions may offer greater design flexibility when complicated interactions discovered from inverse methods are challenging to realize experimentally.\cite{Lindquist2016B, Jadrich2017, Adorf2018} However, a systematic inverse approach is likely required to find these optimal protocols given the complexity of the design space.

\subsection{Computational modeling for experimental materials design}
\label{sec:experimentconnect}
Though inverse design has been successful \emph{in silico}, connecting these techniques with experiments to realize new materials remains a key challenge. Inverse methods can be applied directly to experiments \cite{Wilken2015}, but this approach is ineffective when the experiments are slow, sensitive, or not amenable to automation. In these cases, computation can be leveraged to rapidly screen materials using inverse techniques, and computational predictions can be verified with experiments. This approach has been applied to find bottlebrush polymers with targeted morphologies \cite{Walsh2019}, an optimal director field for a liquid crystal \cite{Aharoni2018}, and disordered materials with targeted acoustic and photonic properties \cite{Batten2008, Florescu2009, Man2013}. The success of the combined computational--experimental approach hinges on (1) the availability of fast and accurate computational models and (2) the ability to constrain the design to experimentally controllable, feasible parameters. The design of block copolymers has given particularly successful demonstrations of this \cite{Hannon2013, Hannon2013B,Qin2013,Khaira2014, Jaeger2016,Paradiso2016, Khadikar2017}, where well-established techniques such as self-consistent field theory can be coupled to inverse schemes.

For other classes of soft materials, reliable models that can access the appropriate length and time scale for assembly may not exist or may be challenging to connect to experiments. Detailed models, where the designable parameters are usually clearer, are often too computationally demanding to use in inverse schemes, while models with effective interactions may not be readily mapped onto experiments. Implementing coarse-graining techniques within inverse frameworks may help bridge this disconnect. One such strategy is to define the design space in terms of experimentally controllable parameters of a detailed model, including parameter constraints. Systematic coarse-graining can then be used to map these parameters to a simpler model for simulating assembly.  For example, for the current value of the design parameters, a fully atomic simulation may be used to compute the effective pair potential between two colloidal particles.  The coarse-grained potential can then be used to simulate assembly of a much larger system of many particles.  The design parameters would then be updated directly and the process iterated. Such an integrated coarse-graining--inverse scheme has not yet been demonstrated but is potentially powerful for connecting computation and experiments to design new materials.

\section{Conclusion}
Inverse approaches suggest systematic means for designing soft materials with complex target structures and desired macroscopic properties.  In this Perspective, we reviewed the methodological and computational challenges associated with various design problems in soft matter and the strategies developed to address them.  Methods for assembling target structures focus mainly on determining an optimal FOM that is descriptive of the target and preferentially encourages its assembly.  Metrics based on thermodynamic energies, statistical distance measures, and structural OPs have all been implemented as FOMs to design interactions that successfully self-assemble a variety of phases with complex structures. These methods may further benefit from ML strategies to automatically discover structural FOMs. For design problems of materials with target properties, the FOM is typically more obvious (i.e., the property itself), so effective strategies focus on developing efficient strategies for determining the relevant structure--property relationships. Since such relations are often computationally demanding to compute, they may benefit from ML strategies to accelerate property evaluation. The advances presented here expand the scope of application for computational design of soft materials and open up promising new opportunities, including the synthesis of reconfigurable materials with multiple functionalities, the engineering of nonequilibrium assembly protocols, and the strengthening of connections between computational and experimental approaches to material discovery.

\begin{acknowledgments}
MPH acknowledges support from the Center for Materials for Water and Energy Systems, an Energy Frontier Research Center funded by the U.S. Department of Energy, Office of Science, Basic Energy Sciences under Award \#DE-SC0019272. BAL acknowledges support from the Darleane Christian Hoffman Distinguished Postdoctoral Fellowship at Los Alamos National Laboratory. TMT acknowledges support of the Robert A. Welch Foundation (Grant No. F-1696).
\end{acknowledgments}

\section*{Data Availability}

Data sharing is not applicable to this article as no new data were created or analyzed in this study.

\bibliography{References}

\end{document}